\documentclass[9pt,twocolumn,twoside]{pnas-new}

\templatetype{pnasresearcharticle} 
\setboolean{displaywatermark}{false}
\title{Thermalization and Possible Signatures of Quantum Chaos in Complex Crystalline Materials}

\usepackage{graphicx}
\usepackage{epstopdf}
\usepackage{amssymb}
\usepackage{array}
\usepackage{amsmath}
\usepackage{color}
\usepackage{bm}
\usepackage{tabularx}
\usepackage[normalem]{ulem}

\author[a,b]{Jiecheng Zhang}
\author[a,b]{Erik D. Kountz}
\author[c]{Kamran Behnia}
\author[a,b,d,1]{Aharon Kapitulnik}

\affil[a]{Department of Physics, Stanford University, Stanford, CA 94305, USA.}
\affil[b]{Geballe Laboratory for Advanced Materials, Stanford University, Stanford, CA 94305, USA.}
\affil[c]{Laboratoire Physique et Etude de Mat\'eriaux (CNRS-Sorbonne Universit\'e), ESPCI Paris, PSL Research University, 75005 Paris, France}
\affil[d]{Department of Applied Physics, Stanford University, Stanford, CA 94305, USA.}

\leadauthor{Kapitulnik} 

\significancestatement{Quantum chaos has been suggested as a framework to understand transport of charge excitations in strongly correlated electron systems exhibiting `strange metal' state without long-lived quasiparticle excitations. Identifying a characteristic local scrambling time and a velocity by which perturbation propagate into non-local degrees of freedom, a chaos diffusivity is identified and related to the measured charge and energy diffusivities. Here we scrutinize the underlying hypothesis that lattice dynamics can be ignored, particularly at high temperatures, where many phonon bands and their interactions dominate the thermal transport in complex materials. We argue that much of the chaotic behavior, which is also identified in complex insulators, originate from phonons, and in equivalent itinerant systems only the thermal (energy) diffusivity describes chaos diffusivity.}

\authorcontributions{Author contributions: J. Z, E.D.K, K.B. and A.K. collected data and performed analyses; J.Z., K.B. and A.K wrote the paper.}
\authordeclaration{The authors declare no conflict of interest.}
\correspondingauthor{\textsuperscript{1}To whom correspondence should be addressed. E-mail: aharonk@stanford.edu}


\begin{abstract}
Analyses of thermal diffusivity data on complex insulators and on strongly correlated electron systems hosted in similar complex crystal structures suggest that quantum chaos is a good description for thermalization processes in these systems, particularly in the high temperature regime where the many phonon bands and their interactions dominate the thermal transport.  Here we observe that for these systems diffusive thermal transport is controlled by a universal Planckian time scale $\tau\sim \hbar/k_BT$, and a unique velocity $v_E$. Specifically, $v_E \approx v_{ph} $ for complex insulators, and $v_{ph} \lesssim v_E \ll v_{F}$ in the presence of strongly correlated itinerant electrons ($v_{ph}$ and $v_F$ are the phonons and electrons velocities respectively). For the complex correlated electron systems we further show that charge diffusivity, while also reaching the Planckian relaxation bound,  is largely dominated by the Fermi velocity of the electrons, hence suggesting that it is only the thermal (energy) diffusivity that describes chaos diffusivity.
\end{abstract}

\begin{document}


\maketitle
\thispagestyle{firststyle}
\ifthenelse{\boolean{shortarticle}}{\ifthenelse{\boolean{singlecolumn}}{\abscontentformatted}{\abscontent}}{}

\section*{Introduction}
It has been recently proposed that thermalization processes in strongly correlated electron systems without well defined quasiparticles exhibit chaotic dynamics. Applied to several model systems, the important ingredients related to quantum chaos, the scrambling time and the speed of its spread \cite{Shenker2014,Roberts2015,Maldacena2016}, could be identified in models of correlated electron systems \cite{Hartnoll2015,Blake2016,Hartnoll2016,Patel1844,Aavishkar2017}. Often, the theoretical scrambling (Lyapunov) time is bound at $\hbar/2\pi k_BT$, suggesting a strong link to a Planckian bound $\tau_p=\hbar/k_BT$, that has been argued to limit  transport diffusion \cite{Blake2016,Hartnoll2016,Aleiner2016,Aavishkar2017,Hartman2017,Patel1844,Davison2017}. While this could be a reasonable approach at very low temperatures, it is clear that at high temperatures such as in the vicinity of Mott-Ioffe-Regel (MIR, most commonly defined as $k_F\ell \sim 1$), one cannot ignore phonons as a natural route for thermalization, particularly in  complex crystal structures where numerous phonon bands span a wide range of temperatures, and multi-phonon scattering processes, enabled by strong anharmonicity dominate thermal relaxation.

Of particular relevance is the recent observation that thermal diffusivity of insulators exhibits a lower bound set by the product of the square of sound velocity and the Planckian time \cite{Behnia2019}. This bound is closely approached for insulators with complex crystal structures, making it conceivable that the phonon system itself approaches thermal equilibrium via chaotic dynamics. For example, a study of spectral statistics of lattice modes in a crystal with a complex unit cell revealed a Wigner-Dyson statistics \cite{Wigner1955,Dyson1962} for the correlations of eigenmode frequencies, a hallmark of chaotic dynamics \cite{Fagas1999,Fagas2000}. Assuming that chaos description applies and since thermal equilibrium is approached through energy eigenstate thermalization \cite{Srednicki1994,DAlessio2016}, it suggests that energy diffusion can be related to chaos diffusion through the characteristic relaxation time and velocity. This could be achieved through the study of thermal diffusivity, especially if the velocity and scattering time could be determined independently \cite{Zhang5378,Martelli2018,Zhang2018}.

In this paper we argue that a large class of complex insulators, as well as bad metals embedded in such insulators, are possible candidates to be analyzed in the framework of a many-body quantum chaos at high temperatures (e.g. in the `bad metal' regime \cite{EmeryKivelson1995}).  Our assertion is based on analyses of thermal diffusivity data on such systems, attributing the origin of chaotic dynamics to the numerous phonon bands that span the measurements temperature range.  Since all material-dependent parameters are known in these solids, we are able to compare their energy (thermal) and charge (electrical) diffusivities. Our primary results are: \textbf{{\it i})} it is highly plausible that quantum chaos is a good description of complex insulators in the high temperature regime where the many phonon bands and their interactions dominate the thermal transport. In particular, a Planckian relaxation time $\tau_p\sim\hbar/k_BT$ and unique velocity $v_E\sim v_{ph}$ ($v_{ph}$ is a phonon velocity of order the speed of sound), characterize the thermal diffusivity in these materials (also see \cite{Behnia2019}); \textbf{{\it ii})} Bad metallic states with no well-defined quasiparticles, that emerge in such insulators upon doping, share much of the dominance of the anomalous phonon thermal transport. However, while the  relaxation time reaches $\tau_p\sim\hbar/k_BT$, the velocity which characterizes the diffusive thermal transport exhibits $v_E \gtrsim v_{ph} $ signifying the appreciable electrons contribution (at the same time, $v_E$ is smaller than the Fermi velocity, $v_F$.); \textbf{{\it iii})} Following a theoretical approach first introduced by Werman {\it et al.}  \cite{Werman2017}, we argue that for these systems we can identify in a consistent way both, a Lyapunov exponent $\Gamma_L\sim 1/\tau$ and a butterfly velocity $v_B \sim v_E$ that take into account the electrons, the phonons, and their interactions; \textbf{{\it iv})} Comparing resistivity with thermal diffusivity, we show that charge diffusivity, while also reaching the relaxation bound,  is largely dominated by the Fermi velocity of the electrons, suggesting that it is the thermal (energy) diffusivity that describes chaos diffusivity.

\section*{Observations and Analyses}

\subsection*{Overview}
Recently, in a comprehensive study of low thermal conductivity insulators, Behnia and Kapitulnik \cite{Behnia2019} analyzed the magnitude of the thermal diffusion constant in the high-temperature regime where T-linear thermal resistivity is observed. Possibly the most striking result of this study has been the observation that complex oxides (such as cubic perovskites), exhibit a thermal relaxation time that is of order (but bound from below) of the Planckian relaxation time $\tau_p=\hbar/k_BT$. More structurally complex insulators such as glasses may come even closer to that bound. Thus, following these observations, we will consider \emph{real} complex material systems such as perovskites where at high temperatures (of order of room temperature or above), a large number of phonon bands are active within each Brillouin zone, many of them dispersive, hence able to carry entropy. Moreover, at the high temperature range of the experiments anharmonic effects are appreciable, thus strongly influencing the many-body states of the phonon system. It is therefore reasonable to assume that the phonon system creates a chaotic environment, exhibiting levels statistics that follows Wigner-Dyson statistics \cite{Fagas1999,Fagas2000}. If in addition these systems are doped, only a very small number of electron bands (often only one) cross the Fermi energy to become relevant. This large number of phonon modes in the presence of appreciable electron-phonon interaction, cause the electrons to locally relax their energy and momentum.

\subsection*{Complex Insulators} 

As alluded above,  even without the contribution of electrons, insulating perovskites exhibit anomalous diffusivity that is likely to be  described as chaotic. 
A ubiquitous $D_{ph}\propto 1/T$ has been highlighted in recent studies of high-temperatures thermal diffusivity of complex insulators, particularly perovskites \cite{Hofmeister2010,Martelli2018}.  In general, such a temperature dependence has been attributed to umklapp scattering of phonons with a scattering rate that decreases as $\theta_D/T$ \cite{Berman1976}, commencing at temperatures as low as $\theta_D/5$ (here $\theta_D=\hbar\omega_D/k_B$ is the Debye temperature  \cite{Slack1979}. However, for the complex oxides, a $\sim1/T$ behavior must have a different origin, since in the relevant temperature range, typically  250K $ \lesssim T\lesssim $ 700K, the phonon mean free path is very small, of order or even smaller than the lattice constant. Therefore, it was often suggested that umklapp scattering alone is unlikely to be sufficiently strong to account for the observed small mean-free path \cite{Vandersande1986,Langenberg2016,Behnia2019}. Furthermore, due to their structural complexity, complex oxides are susceptible to a variety of local disorder effects, which may further enhance relaxation processes and thus degrade thermal transport. These observations lead to a suggestion that the phonon thermal diffusivity is limited by a Planckian relaxation time $\tau \sim \hbar /k_BT$ \cite{Zhang5378,Martelli2018,Behnia2019}. This proposal may first seem at odds with the fact that  the temperature range where this behavior occurs is of order, or exceeding the acoustic phonons Debye temperature, thus should be considered as ``classical.''  However, a closer investigation of the phonons band structure of perovskites and similar complex materials show that new phonon bands, many of them dispersive, continue to be activated much above 1000K, where much of the decrease in phonon mean free path comes from anharmonic and phonon-phonon scattering. For example, MgSiO$_3$ \cite{Wehinger2016} or SrTiO$_3$ \cite{Trautmann2004} exhibit phonon bands all the way to $\sim 1500$K, while for YBa$_2$Cu$_3$O$_7$ phonon bands exceed $\sim 1100$K \cite{Nozaki1993}. By contrast, material systems such as PbTe, exhibit very low thermal diffusivity due to their low sound velocity, but are further away from $\tau_p$ \cite{Behnia2019}.

Table~\ref{perovskites} below is a compilation of the linear behavior parameters of the inverse diffusivity of several examples of insulating perovskites as compared to well known highly crystalline ``simple'' insulators: BeO, silicon and diamond. This table is constructed from data in the range of 250K$ \lesssim T\lesssim $ 700K by assuming a phonon-only thermal diffusivity behavior of 
\begin{equation}
D_E =D_{ph}=\tfrac{1}{d}v_{ph}^2\tau =s \frac{\hbar v_{ph}^2}{k_BT}
\label{phonondiff}
\end{equation}
here $d$ is the relevant dimensionality in the standard expression for the diffusivity, but it is embedded in the numerical constant $s$ once we impose a Planckian relaxation time $\tau \sim \hbar /k_BT$. Furthermore in this expression, $v_{ph}$ is taken as the compressional sound velocity with the rational that much of the heat is transported by the longitudinal acoustic (LA) mode, since it involves excursions of atoms along the direction of heat propagation (similar to \cite{Martelli2018}). Also we add in Table~\ref{perovskites} an estimate of the phonon mean-free path, calculated at room temperature from the expression $D_{ph}\equiv v_{ph}\ell_{ph}$. 

\begin{table}[ht]
\begin{center}
	\caption{Thermal diffusivity parameters for complex insulators, including insulating perovskites, Gadolinium Gallium Garnet (GGG) and for comparison also highly crystalline ``simple'' insulators (references given next to sample description). Using known compressional speed of sound for $v_{ph}$, $s$ is determined from: $D_{ph}=s \hbar v_{ph}^2/k_BT$. The room-temperature phonon mean free path is calculated from $\ell_{ph}\equiv D_{ph}/v_{ph}$. Note that for BeO, Si, and Diamond, linear inverse-diffusivity is observed only above $\sim 750$K, in which regime $s$ is calculated, see Fig.~\ref{plot}. }	\label{perovskites}
\begin{tabular}{|c|c|c|c|}
		\hline\hline
		&$v_{ph}$	&	&$\ell_{ph}({\rm 300 K})$\\
		sample	&[$10^5$cm/s]	&$s$& \AA\\
		\hline
	SrTiO$_3$ \cite{Martelli2018}	&7.87	&2.7&5.1\\
		\hline
	LaAlO$_3$ \cite{Hofmeister2010,Carpenter2010}	&6.72	&2.9&3.86\\ 
		\hline
	KTaO$_3$ \cite{Hofmeister2010,Barrett1968}	&7.5 	&3.1&5.56\\ 
		\hline
	KNbO$_3$	\cite{Hofmeister2010,Kalinichev1993}&7.0 	&1.6&2.69\\ 
		\hline
	NdGaO$_3$	\cite{Hofmeister2010,Krivchikov2000}&6.5 	&1.65&2.61\\ 	
		\hline
	YAlO$_3$		\cite{Hofmeister2010,Zhan2012}&8.25 	&1.8&3.54\\ 	
		\hline
	MgSiO$_3$	\cite{Osako1991,Kung2004}	&8.0 	&1.05&2.1\\ 	
		\hline
	disordered-SrTiO$_3$ \cite{Hofmeister2010} &7.87&1.9&3.57\\ 
		\hline
	GGG		\cite{Fan2007,Kitaeva1985}	&6.55 	&2.5&3.98\\ 	
		\hline	
	PbWO$_4$	\cite{Cai2011,Kavitha2015} 	&3.47 	&3.0&2.65\\ 	
	 \hline   \hline\hline
	BeO	\cite{Hofmeister2014,Kiiko2007}&11.3 	&11&46\\ 	
		\hline
	Silicon	\cite{Martelli2018}	&8.43 	&23&202\\ 
		\hline		
	Natural-diamond	\cite{Victor1962,Olson1993,Wang2004}	&18.0 	&50&55000\\ 						
		\hline\hline
	\end{tabular}
	\end{center}
\end{table}

The most important point to note from this table is that complex oxides are very different than the simple highly crystalline insulators even though the high-temperature diffusivity in both cases is inversely proportional to temperature. The observation that they saturate the thermal relaxation bound at the Planckian rate, while at the same time exhibit numerous phonon bands that are likely obeying Wigner-Dyson statistics \cite{Fagas1999,Fagas2000} suggest that they satisfy the conditions to exhibit chaos and quantum thermalization \cite{Srednicki1994}. This is an important observation, as we next superimpose the electronic contribution to the thermal transport.\\

\subsection*{Bad metals embedded in complex insulators}  
In the presence of electrons, phonon-electron scattering rate may be  the dominating high-temperature cause of the observed short phonon mean free path in complex correlated materials (see e.g.  \cite{Steigmeier1964,Vandersande1986}). Indeed, in two recent studies of hole doped \cite{Zhang5378} and electron doped \cite{Zhang2018} cuprates, Zhang {\it et al.} have shown that the high temperatures thermal diffusivity of these complex oxides exhibit electronic as much as phononic character. These perovskite material systems are known to  have many phonon bands that are excited at temperatures above the resistively-determined MIR limit, while also possessing substantial electron-phonon interaction. However, the main discovery of these two papers has been that over a very wide range of thermal-modulation frequency the data could be fit with a single diffusivity constant of $D_E$ corresponding to the interacting electron-phonon system.  Moreover, analyzing the temperature dependence of the thermal diffusivity, Zhang {\it et al.} noticed that in the case of electron-doped cuprates (L$_{2-x}$Ce$_x$CuO$_4$, with L=Nd, Sm and Pr), the measured inverse thermal diffusivities at high temperatures are linear in temperature, satisfying:
\begin{equation}
	D_E^{-1}=\Big(s\frac{\hbar v_E^2}{k_BT}\Big)^{-1}+D_0^{-1}
	\label{linear}
\end{equation}
where $s$ plays the same role as in the pure phonon case. The T-linearity is directly attributed to a Plankian relaxation rate $\tau=\hbar/k_BT$, and the slope of the linear term gives rise to a velocity $v_E$, which was interpreted as the velocity of an incoherent mixture (`soup') of electrons and phonons. The difference between Eqn.~\ref{linear} and Eqn.~\ref{phonondiff} is the $T\to 0$ extrapolation which is close to zero for the insulators, but finite for the systems that include electrons. The constant term, $D_0^{-1}={\rm lim}_{T\to 0}D_E^{-1}(T)=(\hbar/3m^*)^{-1}$ was interpreted as the residual electron diffusivity contribution at the MIR limit  \cite{Zhang2018}.  The presence of electrons must imply that either $s$, or the velocity, or both need to be modified to include the electrons. There are three possible scenarios for in this case. {\it i}) we assume that the velocity remains the sound velocity. In that case we can show that in the range of materials we measured $s$ must increases 4 to 6 times from its present value, implying an increasing mean free path. This does not make sense since the presence of electrons tend to increase the phonon relaxation through electron-phonon interaction, hence is expected to reduce $s$. Furthermore, from our YBCO study we know that the thermal diffusivity follow the resistivity and not the sound-velocity anisotropy, thus should have implicit character of the electron contribution.{\it ii})  we can assume that the electrons indeed push the relaxation time to the Planckian limit with $s=1$. While this is certainly possible, and would yield $v_E$ much larger than $v_s$, we choose to be more conservative and choose {\it iii}) we assume that 
 since the system is dominated by phonons, similar to the parent insulator,  only the velocity is changing, taking $s \approx 2.0$ calculated as an average value of many different perovskites (e.g. the average of the systems in Table~\ref{perovskites}).  Thus, determining $s$,  $v_E$ can be calculated using Eqn.~\ref{linear}. Table~\ref{fittable} is a summary of results for a few materials reported in that study \cite{Zhang2018}.
\begin{table}[ht]
\begin{center}
\caption{Velocity $v_E$ extracted from the linear fit to Eqn.~\ref{linear} of measured thermal diffusivity for L$_{2-x}$Ce$_x$CuO$_4$, with L = Nd (N), Sm (S), and Pr (P)  and BSCCO data taken from ref.~\cite{Zhang2018} ((g) is for as-grown, while  (n) is for annealed samples), as well as YBCO data from \cite{Zhang5378} (here (a) and (b) represent the plane and chain directions respectively). Also listed are the measured speed of sound \cite{Fil1996,Berggold2006} and Fermi velocity \cite{Armitage2010}. The room-temperature  mean free path is calculated from $\ell_{eff}\equiv D_E/v_{E}$. }
	\label{fittable}
\begin{tabular}{|c|c|c|c|c|c|c|}
		\hline\hline
				&$v_s$		&$v_E(s=2)$	&$v_F$			&$\ell_{eff}({\rm 300K})$		\\
		sample	&[$10^5$cm/s]	&[$10^5$cm/s]	&[$10^7$cm/s]	& [\AA ]\\
		\hline
		NCCO$_{0.15}$(g)	&$7.0$	&$12.0$	&$2.5$	&3.4	\\
		\hline
		NCCO$_{0.15}$(n)	&$7.0$	&$11.3$	&$2.5$	&3.6	\\ 
		\hline
		SCCO$_{0.16}$(g)	&$5.9$ 	&$12.0$	&$2.0$	&2.8	\\ 
		\hline
		SCCO$_{0.16}$(n)	&$5.9$ 	&$10.6$	&$2.0$	&3.0	\\ 		
		\hline
		PCCO$_{0.13}$(g)	&$6.25$	&$12.7$ &$2.1$	&3.3	\\ 	
		\hline
		BSCCO 				&$4.37$ 	&$12.0$	&$2.4$	&1.75		\\ 		
		\hline
		YBCO$_{6.75}$(a) 	&$6.05$ &$7.8$	&$2.25$	&2.2	\\ 	
		\hline
		YBCO$_{6.75}$(b) 	&$6.5$ 	&$10.6$	&$-$	&3.4	\\ 	
		\hline
		YBCO$_{6.60}$(a) 	&$6.05$ &$7.8$	&$2.1$	&2.4	\\ 	
		\hline
		YBCO$_{6.60}$(b) 	&$6.5$ 	&$10.0$	&$-$	&4.0	\\ 	
		\hline\hline
	\end{tabular}
	\end{center}
\end{table}

The most important feature in Table~\ref{fittable} is that $v_E \gtrsim v_{ph}$ by roughly a factor of two, which translates to a factor of four in the diffusivity slope (although, as discussed above, the difference can be larger if $s$ is closer to unity). 
Furthermore, note that the extracted $\ell_{eff}$ in that table is larger than for the pure insulators summarized in Table~\ref{perovskites}. For example, NdGaO$_3$ exhibits similar sound velocity, but much smaller extracted phonon mean free path than any of the LCCO samples. This can be understood as a result of the contribution of the electrons to the thermal transport, which is another reason to argue for their role in energy transport even at these high temperatures.

\begin{figure}[ht]
	\centering
	\includegraphics[width=1.0\columnwidth]{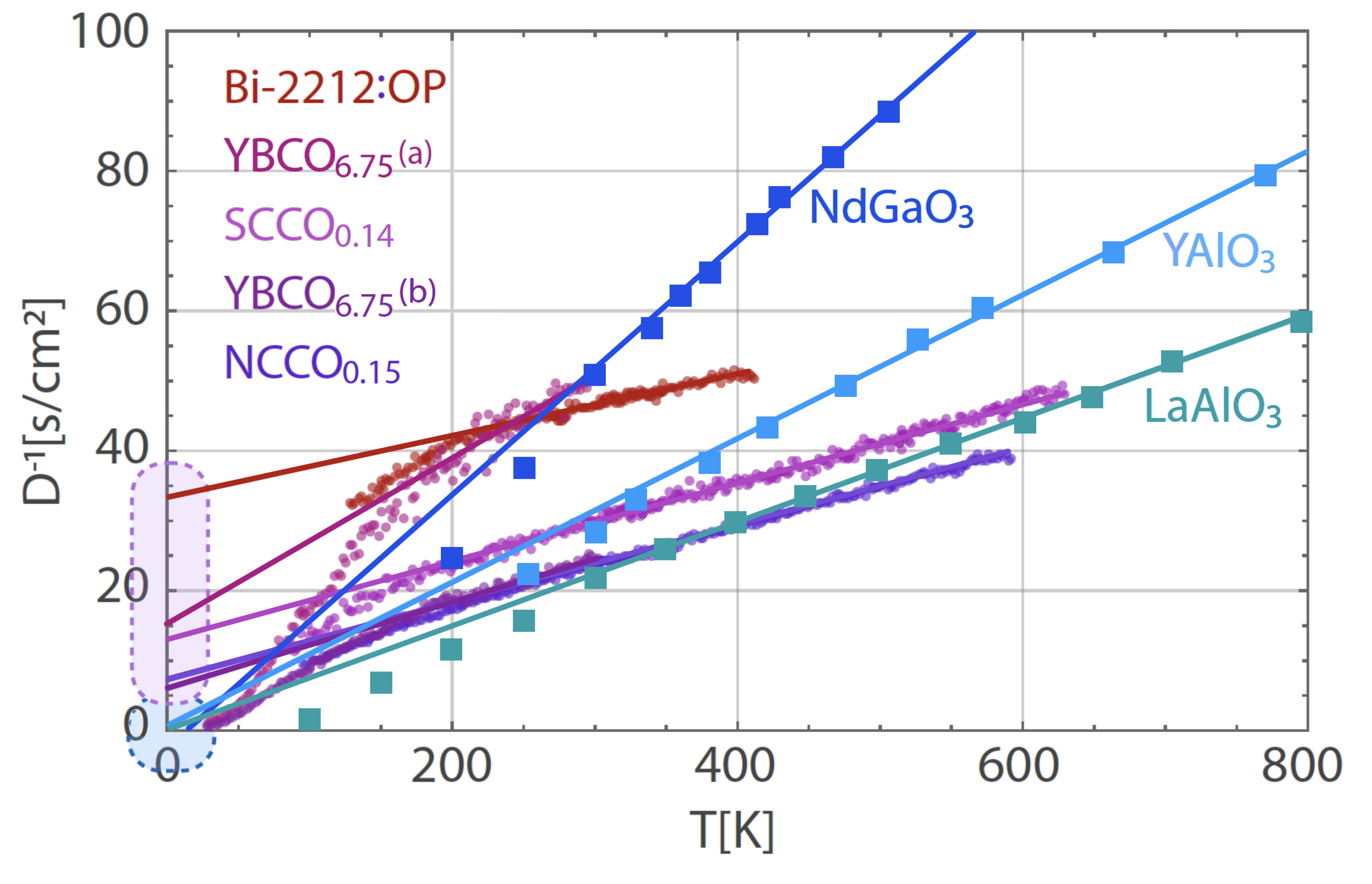}
	\caption{Inverse thermal diffusivity data plotted vs. temperature for several complex materials. Diffusivities of insulating LaAlO$_3$, YAlO$_3$ and NdGaO$_3$ are calculated from \cite{Schnelle2001,Hofmeister2010}. Diffusivities of complex `bad metals' are taken from earlier publications: Bi-2212, SCCO and NCCO \cite{Zhang2018}, and YBCO \cite{ Zhang5378}.}
		\label{plot}
\end{figure}

To end-up the discussion on thermal diffusivity, we show in Fig.~\ref{plot} a log-log plot of the diffusivity vs. temperature above 100 K for several insulating samples as compared to Nd$_{1.85}$Ce$_{0.15}$CuO$_4$ (NCCO) . First we note that silicon, as a highly crystalline material approaches $D_E \sim T^{-1}$ at high temperatures, but this can be attributed to standard umklapp scattering \cite{Glassbrenner1964}. On the other hand, YAlO$_3$ and NdGaO$_3$ are complex insulating materials with non-cubic crystal structure similar to NCCO, with diffusivity that follows Eqn.~\ref{phonondiff}, and with a typical proportionality constant $s$ (See table \ref{perovskites}). The NCCO crystal which is also a complex perovskite with a very similar sound velocity shows also linear dependence of the inverse diffusivity, but with a finite intercept. Subtracting that intercept, its linear  dependence on the same log-log plot is evidence. Also evident is the higher slope, by about a factor of 2 to 3 over the insulators. Keeping the prefactor constant (see above), this is a manifestation of the higher velocity extracted from the slope.  To emphasize the range of phonon bands in these materials, we note e.g. that for Nd$_2$CuO$_4$ \cite{Pintschovius1991} (the parent compound of NCCO), or YAlO$_3$ \cite{Suda2003}, room temperature lies low within the rather wide (extending much above 1000K) complex band structure of the phonons, some of which are strongly dispersive.\\

\subsection*{Charge Diffusivity and Resistivity} 
In the presence of itinerant electrons, such as in the aforementioned strongly correlated systems, we can discuss separately the issue of charge transport and its relation to the MIR limit. For the strongly correlated hole or electron doped cuprates studied by Zhang {\it et al.} \cite{Zhang5378,Zhang2018}, it was argued that both, charge and lattice degrees of freedom are strongly scattered, hence the scattering rate is pushed to the Planckian bound \cite{Zaanen2004}. Applying a temperature gradient to the sample, that system of electrons and phonons react together, and heat is flowing as an  overdamped diffusing `soup' of phonons and electrons. While the very high density of phonons dominate the thermal transport, the electrons which with higher velocity but the same scattering rate are able to increase the effective velocity of that fluid.

Considering next charge transport, an electric field applied to the system will cause charges to move, but will only influence the phonon system indirectly through electron-phonon interaction. This in turn may only result is some renormalization of the electron's effective mass, but will not change the velocity that determines the charge diffusivity, that is, the Fermi velocity, by much. Therefore, establishing that $v_{ph} \lesssim v_E \ll  v_{F} $, we conclude that in general $D_C  \gtrsim  D_E$. 

To further explore the above idea, we compare resistivity and thermal diffusivity data, particularly on materials that show $D_E^{-1}\propto T$. Using Einstein relation for the resistivity, the charge diffusivity is given by $D_C=[e^2(dn/d\mu)\rho]^{-1}$. If the relaxation time is bound at $\tau \sim \hbar/k_BT$, the resistivity will show $\rho \propto T$, and if the density of states is known, the velocity associated with the charge diffusivity can be calculated. It is evident from the resistivity data compiled by Bruin {\it et al.} \cite{Bruin2013} that the extracted velocity associated with $D_C^{-1}$ is approximately the Fermi velocity. One concrete example is Bi$_2$Sr$_2$CaCu$_2$O$_8$ (BSCCO), which is a hole-doped cuprated that is less susceptible to oxygen loss and thus was measured to temperatures much above room temperature. In particular, for near optimal doping a linear dependence of the resistance and inverse thermal diffusivity have been observed. While the thermal diffusivity yields a velocity $v_E = 7.4 \times 10^5$ cm/s, analysis of the linear resistivity on very similar crystals (see e.g. \cite{Vedeneev2005}) yields a velocity very close to the known Fermi velocity for this system $v_F \sim 2.8\times 10^7$ cm/s \cite{Vishik2010}.

A more striking difference between energy and charge diffusions is observed in many of the electron doped cuprates where above the MIR limit the inverse thermal diffusivity is linear in temperature \cite{Zhang2018}, while the resistivity behaves as $\rho\propto T^{1+y}$, with $y\geq 0$  \cite{Bach2011,Scanderbeg2016,Sarkar2018}. Imposing a Planckian relaxation bound suggest that in these materials the electronic compressibility varies as well above the MIR limit as has been recently proposed in \cite{Perepelitsky2016,Werman2016}.\\

\section*{ Discussion} 
In a recent publication Werman {\it et al.} \cite{Werman2017} introduced the idea that a strongly coupled incoherent bad metals are also strongly chaotic. In their model an electron system interacts with a parametrically much larger system of optical phonons via local electron-phonon interaction. The root-mean-squared phonon frequency and velocity averaged over all phonon bands were set to be $\omega_0$ and $v_{ph}$ respectively, while similarly, the root-mean-squared Fermi  velocity averaged over all electron bands was set to be $v_{el}$. Their calculations, which were performed for the cases of dispersive and dispersionless phonons, concentrated on the temperature regime where $\omega_0 \ll k_BT \ll \epsilon_F$, in which electronic quasiparticles are no longer well-defined. With the above assumptions, their primary result was that the thermal and chaos diffusion constants, $D_E$ and $D_\chi$  are always comparable ($D_E\sim D_\chi$), while this is not necessarily the case for charge diffusion. More recently similar results were also obtained by Guo {\it et al.} \cite{Guo2019}, who introduced to the aforementioned SYK model \cite{Davison2017}, in addition to the electronic coupling of the SYK islands, also a local coupling of each island to a large number of low-energy ``well defined'' phonons. 

\subsection*{Quantum Chaos and Thermalization} 
\begin{figure}[ht]
	\centering
	\includegraphics[width=1.0\columnwidth]{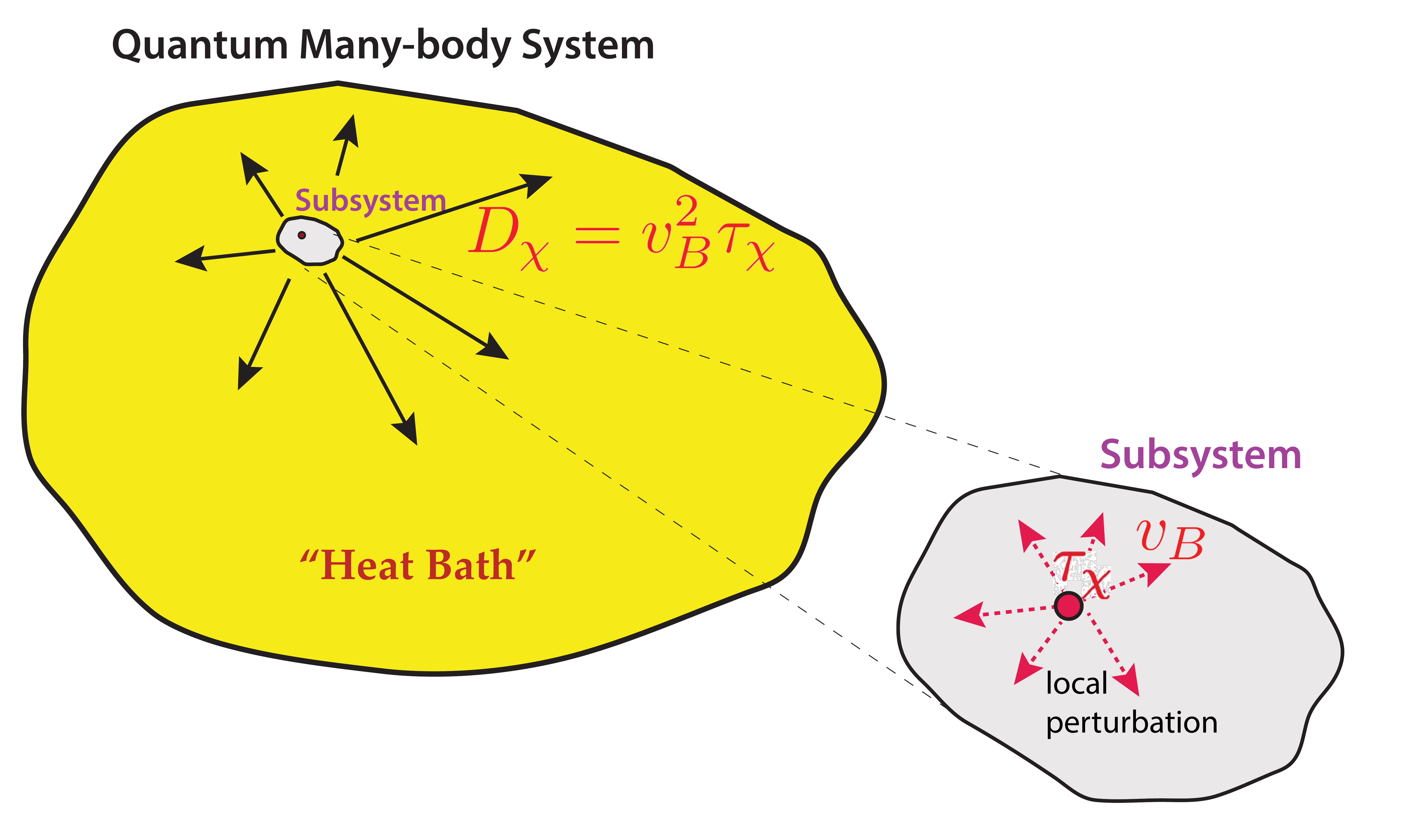}
	\caption{Cartoon showing the initial local perturbation that leads to scrambling in the subsystem followed by global thermalization. }
		\label{thermalization}
\end{figure}

A closed, many-body interacting quantum system without well defined quasiparticle excitations is argued to exhibit quantum chaos (see e.g. \cite{Hartnoll2016}). In this scenario local perturbations in the system's initial conditions are ``scrambled'' exponentially into non-local degrees of freedom with a Lyapunov exponent rate $\Gamma_\chi = 1/\tau_\chi$, and spread out to affect the whole system at the butterfly velocity $v_B$ \cite{Shenker2014,Blake2016,Hartnoll2016,Hartman2017,Patel1844}. Consider a perturbed subsystem of a larger many-body interacting quantum system. It is then possible to assign a thermal distribution to the subsystem, while the rest of the system serves as a heat bath that allows the subsystem to thermalize. Scrambling and the butterfly velocity then describes the initial amplification of local perturbations and the spread of chaos through quantum entanglement and the loss of memory of the initial state of the whole system. Macroscopic diffusion originated from the microscopic scrambling processes emerges, with a diffusion constant
\begin{equation}
D_\chi \approx v_B^2\tau_\chi \ ,
\end{equation}
that controls global thermalization. Figure~\ref{thermalization} shows a cartoon of this sequence, which was further argued to describe charge and energy diffusion \cite{Blake2016,Hartnoll2016,Aavishkar2017}.

\subsection*{Quantum Chaos and Solid State Systems} 
While it is reasonable to expect quantum chaos in strongly interacting electron systems, its observation, as well as that of a scrambling time, are generally not expected to be measurable in solid state systems. However, it was recently proposed that either the charge diffusivity $D_C$ \cite{Blake2016,Blake2016a} or energy diffusivity $D_E$ \cite{Patel1844}, or both \cite{Werman2017}, are related to $D_\chi$. These connections followed a proposal by Hartnoll \cite{Hartnoll2015}, that for incoherent non-quasiparticle transport, both diffusivities are subject to fundamental quantum mechanical bound of $D \gtrsim \frac{\hbar v_F^2}{k_BT}$,  where the Fermi velocity was introduced as the natural velocity for the problem of electronic transport in a metal, and the relaxation time was set at the shortest time scale allowed by the uncertainty principle $\tau \sim \hbar/k_BT$ \cite{subirbook}. An immediate consequence of this approach is that for a constant  electronic compressibility, the resistivity is linear in temperature, which has been ubiquitously observed in many bad metals \cite{Bruin2013}. 

While the above proposal seems to work for charge diffusivity ($D_C$), it was further suggested that the above bound applies to energy diffusivity $D_E$ and that $D_E\sim D_C \sim D$, with the proviso that electron-phonon interaction can be neglected. However, in calculations of charge and energy diffusivities in specific models of metals without quasiparticles, a critical Fermi surface \cite{Patel1844}, and the Sachdev-Ye-Kitaev (SYK) model \cite{Davison2017}, it was found that ``quantum chaos, as characterized by the butterfly velocity and the Lyapunov rate, universally determines the thermal diffusivity, but not the charge diffusivity''  \cite{Hartnoll2016,Davison2017}. This seems natural in the context of the studied models since quantum chaos originates from local energy fluctuations.

Obviously, an interacting many-electron system in an itinerant solid is not in general a closed quantum system, since at least phonons need to be incorporated into the problem. Moreover, any experiment on a solid state system in which the temperature is well defined implies a connection to a heat bath which in general is dominated by the phonons. Thermalizing the system is a delicate balance between internal electron-phonon interaction and connection to the environment through e.g. the electrical leads. Thus, even if we invoke a saturation of the relaxation time at $\hbar/k_BT$, the notion of a unique velocity that governs scrambling of local perturbations into non-local degrees of freedom, that is, the ``spread of chaos'' may not be applicable to solid state systems. However, we can still use the concept of quantum-chaos in situations that fulfill certain conditions. In particular, a state that at least mimics a quantum chaotic system may be realized in a regime where the full electron-phonon system includes electron-phonon interactions, and also exhibits strong  momentum degradation.

\subsection*{Comparison to Experiments}
We return now to the comparison of our experimental observations in perovskites and other bad metals and complex insulators to the model of Werman {\it et al.} \cite{Werman2017}. In that model, strongly renormalized electrons are coupled to weakly interacting optical phonons, which are well defined quasiparticles. This yields an average phonon frequency $\langle \omega_{ph}^2\rangle^{1/2} \equiv \omega_0 \ll k_BT$, and thus $\tau_\chi\sim k_BT/\hbar\omega_0^2$, which is far from the Planckian bound, thus should only be considered as ``weakly chaotic. Scrambling rate is therefore determined by the loss of phonon phase coherence, and  the butterfly velocity of the electron-phonon system with dispersive phonons depends only on the phonons velocity. There are two main differences between the theoretical model and the conditions of the experimental data. First, concerning the phonon system, since the temperature range of interest lies in the middle of the complex phonon band structure, which also exhibits strong phonon-phonon interactions, we cannot assume that all phonons are either well defined quasiparticles, nor we can implement the theoretical assumption of $k_BT \gg \omega_0$. In fact, we argue that the complexity of the phonon system in the studied material systems implies phonon states that most likely exhibit chaotic statistics \cite{Fagas1999,Fagas2000}. The magnitude of the measured diffusivities suggest that much of the heat is transported by dispersive phonon modes, most probably involving excursions of atoms along the direction of heat propagation, and the relaxation reaches  $\tau\sim\hbar/k_BT$. Since in experiments $k_BT$ determines both, the excited phonon modes and relaxation rate,  it is tempting to push the theoretical model to that limit, requiring that the average phonon frequency is determined by temperature, that is $\hbar\omega_0 \to k_BT$,  which immediately implies $\tau_\chi \sim \hbar/k_BT$. This assignment also suggests that with no charge carriers, the velocity associated with the diffusivity must be similar to the phonon velocity, which we summarized in table~\ref{perovskites}, yielding $v_B\sim v_{ph}$. 
 
 Adding the effect of the electrons for the further comparison with ref~ \cite{Werman2017}, the fact that a single diffusion constant is observed suggest an increase in butterfly velocity $v_B\sim v_E \gtrsim v_{ph}$, but probably smaller than $v_F$. This again is understood as a consequence of the strongly interacting electron-phonon system, which diffuses as a `soup' of not well-defined quasiparticles \cite{Zhang5378}, with a diffusion constant that is characterized by a relaxation time saturated at the Planckian time of $\sim \hbar/k_BT$, and a velocity  $v_E>v_{ph}$ due to the contribution of the much faster electrons.  Indeed, Guo {\it et al.} \cite{Guo2019} remark that a more self consistent inclusion of the electron-phonon and phonon-phonon interactions is expected to yield an ``electron-phonon" soup, in which the strong dependence on the electron-phonon coupling disappears.

\subsection*{Final Remarks}
There are two final remarks to make. First, it would be desired to be able to separate the values of $v_E$ and $s$ in Eqn.~\ref{linear}. This may require measurements at shorter timescales to locate the actual time at which electrons thermalize with phonons.  Second, we remark on the relationship  between $D_E$ and $D_C$. For energy diffusion,  both electrons and phonons move along the temperature gradient, and for overwhelming phonon system the electrons are a small contribution to the combined `soup' providing enhanced velocity. On the other hand, an electric field will only act on the charges, which will move close to the Fermi velocity as conjectured in \cite{Hartnoll2015}. The phonons will then work to impede that motion through the interaction with the charges. Therefore, and in particular based on our experimental observations, we \underline{do not} see a reason why  $D_C\sim D_E\sim D_\chi$.

\matmethods{ Data used in this paper was taken from published literature as indicated in the relevant references. Properties used to establish various claims have been calculated as described in the text.


}

\showmatmethods{} 

\acknow{We thank Steve Kivelson, Subir Sachdev, and particularly Sean hartnoll 
for many comments and insightful discussions. This work was supported by the Gordon and Betty Moore Foundation through Emergent Phenomena in Quantum Systems (EPiQS) Initiative Grant GBMF4529, and by the U. S. Department of Energy (DOE) Office of Basic Energy Science, Division of Materials Science and Engineering at Stanford under contract No. DE-AC02-76SF00515. }

\showacknow{} 

\bibliography{Chaos_S}

\end{document}